\documentclass[amsmath,prb,twocolumn,showpacs,floatfix]{revtex4}
\usepackage{graphicx}
\usepackage{amssymb}
\newfont{\Sc}{eusm10}

\begin{document}
\title{Superconducting-normal interface propagation speed in superconducting samples}

\author{Artorix de la Cruz de O\~{n}a}
\email{Artorix.DelaCruz@nbed.nb.ca} \affiliation{A. Center of
theoretical Physics and Applied Mathematics, Dynamical System
Project, Montr\'eal, H3G 1M8, Canada.\\ District Scolaire 9, 3376
rue Principale C.P. 3668, NB E1X 1G5, Canada.}
\date{\today}
%
\vspace{-15mm}
\begin{abstract}
In this paper a new approach to obtain the interface propagation
speed in superconductors by means of a variational method is
introduced. The results of the approach proposed coincide with the
numerical simulations. The hyperbolic differential equations are
introduced as an extension of the model in order to take into
account delay effects in the front propagation due to the pinning.

\end{abstract}
\vspace{5mm}
\maketitle
%
\section{Introduction}
The study of interface propagation is one of the most fundamental
problems in nonequilibrium physics. The understanding of the
magnetic field penetration or expulsion in Superconducting samples
has been a major challenge. An important problem to be solved is the
determination of the speed at which the interface moves from a
superconducting to a normal region.

In Ref.\onlinecite{barto}, Di Bartolo and Dorsey have obtained the
front speed by using heuristic methods such as Marginal stability
hypothesis(MSH) and Reduction order.

In general, the nonlinear equations have been employed to model
fronts propagation in different areas such as population growth and
chemical reactions. Our start point is the nonlinear diffusion
equations(ND) of the form $u_{t}=u_{xx}+f(u)$ obtained from the
Ginzburg-Landau expressions\cite{dorsey}(GL). The GL comprise a
coupled equations for the density of superconducting electrons and
the local magnetic field.

Benguria and Depassier\cite{Bengu1,Bengu2,Bengu3} have developed a
variational speed selection method(BD) to compute the front speed in
ND equations. In the BD method a trail function $g(x)$ is defined
\emph{a priori} and one may find accurate lower and upper bounds for
the speed $c$. Only if the lower and the upper bounds coincide, then
the value of $c$ can be determined without any uncertainty. To
eliminate the ambiguity in the speed determination, Vincent and Fort
in Ref.\onlinecite{vincent} have proposed a more accurate approach
based on the BD method. The approach assumes some approximative
considerations from where the function $g(x)$ is determined.

The purpose of this paper is to develop further the insights into
the front propagation afforded by the work in
Ref.\onlinecite{barto}. We aboard the determination of the
propagation speed from a variational point of view. We use an
alternative approach to the one developed by
Vincent-Fort\cite{vincent}.

In order to describe the evolution of the system between two
homogeneous steady state, we assume a SC sample embedded in a
stationary applied magnetic field equal to the critical $H=H_{c}$.
The magnetic field is rapidly removed, so the unstable
normal-superconducting planar interface propagates toward the normal
phase so as to expel any trapped magnetic flux, leaving the sample
in Meissner state. We have considered that the interface remains
planar during all the process.

The existence of a delay time in the interface propagation systems
is an important aspect that can be modeled by hyperbolic diffusion
equations(HD) which generalize the ND. The HD has been recently
applied in biophysics to model the spread of humans\cite{fort},
bistable systems\cite{mendez1}, forest fires\cite{mendez2} and in
population dynamics\cite{mendez3}. With the goal to take into
account the delay effect on the interface propagation
speed\cite{dorsey} in superconductors, due to, for example,
imperfections, vortex-vortex interactions, the presence of
pinning\cite{altshuler,brandt}, we have included the relaxation time
$\tau$ for the front, and indee introduce the hyperbolic
differential equations.
%

\emph{Traveling wave solutions}. In this paper, we are interested in
the one-dimensional time-dependent Ginzburg-Landau equations, which
in dimensionless units\cite{dorsey} are:
$\partial_{t}f=(1/\kappa^2)\,\,\partial^{2}_{x}f-q^{2}f+f-f^{3}$ and
$\bar{\sigma}\partial_{t}q=\partial^{2}_{x}q-f^{2}q$.

Here, the quantity $f$ is the magnitude of the superconducting order
parameter, $q$ is the gauge-invariant vector potential (such that
$h=\partial_{t}q$ is the magnetic field), $\bar{\sigma}$ is the
dimensionless normal state conductivity (the ratio of the order
parameter diffusion constant to the magnetic field diffusion
constant) and $\kappa$ is a parameter which determines the type of
superconducting material; $\kappa<1/\sqrt{2}$ describes what are
known as type-I superconductors, while $\kappa>1/\sqrt{2}$ describes
what are known as type-II superconductors.

We are interested in finding traveling wave solutions for our model.
We will search for steady traveling waves solutions for the GL
equations of the form $f(x,t)=s(x-c\,t)$ and $q(x,t)=n(x-c\,t)$,
where $z=x-c\,t$ with $c>0$. Then the equations become
\begin{eqnarray}
\label{eq:steady_equation}
\frac{1}{\kappa^2}\,\,s_{zz}+c\,s_{z}-n^{2}s+s-s^{3}=0\nonumber,
\\
n_{zz}+\bar{\sigma}c\,n_{z}-s^{2}n=0,
\end{eqnarray}
\section{Variational analysis}
\emph{Vector potential $q=0$}. In this section, we assume $q=0$ for
the GL equations,
\begin{eqnarray}
\label{eq:newginzburg}
\partial_{t}f=\frac{1}{\kappa^2}\,\,\partial^{2}_{x}f+f-f^{3}.
\end{eqnarray}

Then, there exists a front $f=s(x-ct)$ joining $f=1$, the state
corresponding to the whole superconducting phase to $f=0$ the state
corresponding to the normal phase. Both states may be connected by a
traveling front with speed $c$. The front satisfies the boundary
conditions $\lim_{s\rightarrow-\infty} f=1,\lim_{s\rightarrow\infty}
f=0$. Then Eq.(\ref{eq:newginzburg}) can be written as,
\begin{eqnarray}
\label{eq:monot_equat}
s_{zz}+c\,\kappa^{2}\,s_{z}+\mathfrak{F}_{k}(s)=0,
\end{eqnarray}
where $\mathfrak{F}_{k} = \kappa^{2}\,s(1-s^{2})$ and
$\mathfrak{F}=(1/\kappa^{2})\mathfrak{F}_{k}$.

We define $p(s)=-ds/dz>0$ and $g$ such that $h=-dg/ds>0$. Taking
into account $hp+(g \,\mathfrak{F}_{k}/p)\geq 2\,\sqrt{g\,h
\,\mathfrak{F}_{k}}$ and following the BD method\cite{Bengu1} we
arrive to
\begin{eqnarray}
\label{eq:varia_veloci}
c\geq\,(2/\kappa)\,\int^{1}_{0}
(g\,h\,\mathfrak{F})^{\frac{1}{2}}\,\, ds / \int^{1}_{0} g\,\, ds.
\end{eqnarray}

Now, the asymptotic speed of the front for sufficiently localized
initial conditions may be determined in the limit
$t\rightarrow\infty$. In the limit one has $s\rightarrow1$ for
$z\rightarrow-\infty$, and $s$ is a slowly varying function of $z$.
Therefore one has $s_{zz}\ll\,s_{z}$, and from
Eq.(\ref{eq:monot_equat}) we have that
$\kappa^{2}\,c\,s_{z}+\kappa^{2}\,\mathfrak{F}(s)\simeq0$ and $
p\simeq\,-s_{z}\simeq\,\mathfrak{F}/c$.

Assuming $p=\mathfrak{F}(s)/\alpha>0$, where $\alpha$ is a positive
constant to determine, we can write in general form the trial
function as,
\begin{eqnarray}
\label{eq:trii}
g(s)=\exp\,\left(-\,\alpha^{2}\int\,\mathfrak{F}^{-1}(s)\,ds\right).
\end{eqnarray}

Multiplying in both sides by the function $h$ in the expression
$\mathfrak{F}_{k}g/p=hp$, we have that,
\begin{eqnarray}
\label{eq:byh}
h\mathfrak{F}_{k}g=h^{2}p^{2}
\end{eqnarray}
By using Eq.(\ref{eq:byh}), the relation $h\,\mathfrak{F}_{k}g=
h^{2}\mathfrak{F}^{2}/\alpha^{2}$ is obtained. Then, the following
relation is valid,
\begin{eqnarray}
\label{eq:newsecond}
2\,\sqrt{h\,\mathfrak{F}_{k}\,g}=\,\frac{2}{\alpha}\,h\,\mathfrak{F}.
\end{eqnarray}
Substituting  Eq.(\ref{eq:newsecond}) in Eq.(\ref{eq:varia_veloci}),
the general expression for the speed is given by,
\begin{eqnarray}
\label{eq:secondveloci}
c\simeq\,max_{\alpha\in
(0,1)}\left(\frac{2}{\alpha\,\kappa}\,\int^{1}_{0}
\mathfrak{F}(s)\,h(s)\,\, ds/\int^{1}_{0} g(s)\,\, ds\right).
\end{eqnarray}
\begin{figure}[!tbp]
\centerline{
\includegraphics[width=3.2in]{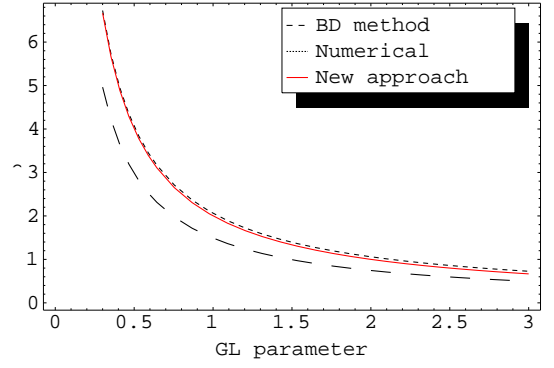}
}\caption{Illustration of the front speed obtained by different
methods in the $q=0$ case versus the GL parameter.}\label{figure1}
\end{figure}

Taking into account the form of $\mathfrak{F}$ and
Eq.(\ref{eq:trii}), the trial function can be written as
\begin{eqnarray}
\label{eq:gtrio} g(s)=\left[(s^{2}-1)/s^{2}\right]^{\alpha^{2}/2},
\end{eqnarray}
and the function $h(s)$,
\begin{eqnarray}
\label{eq:htrio}
h(s)=\alpha^{2}\,s^{-(1+\alpha^{2})}\,(1-s^{2})^{(\alpha^{2}-2)/2}.
\end{eqnarray}
The integrals in Eq.(\ref{eq:secondveloci}) are given by,
\begin{eqnarray}
\label{eq:newinte1}
\int^{1}_{0}g(s)\,ds=\frac{1}{\sqrt{\pi}}\,\,\Gamma\left((1-\alpha^{2})/2\right)\,\Gamma\left(1+\alpha^{2}/2\right),
\end{eqnarray}
which is valid for $\alpha\geq 0$,
\begin{eqnarray}
\label{eq:newinte2}
\int^{1}_{0}f(s)\,h(s)\,ds=\frac{1}{\sqrt{\pi}}\,[\,\Gamma\left((1-\alpha^{2})/2\right)\\\nonumber
-2\,\Gamma\left((3-\alpha^{2})/2\right)]\,\Gamma\left(1+\alpha^{2}/2\right),
\end{eqnarray}
which is valid for $0<\alpha<1$.

Replacing Eqs.(\ref{eq:newinte1}) and (\ref{eq:newinte2}) in
Eq.(\ref{eq:secondveloci}), we arrive to the speed for the front,
\begin{eqnarray}
\label{eq:Otheresultsev}
c\simeq\,max_{\alpha\in
(0,1)}\,\,\frac{2}{\kappa\,\alpha}\,\left(1-\frac{2\,\Gamma\,\left[\frac{1}{2}\,(3-\alpha^{2})\right]}{\Gamma\,\left[\frac{1}{2}\,(1-\alpha^{2})\right]}\right).
\end{eqnarray}
Notice that for $\alpha=1$, we obtain the maximum for
Eq.(\ref{eq:Otheresultsev}), $c=2/\kappa$ which is the result
obtained by using the MSH method.

In Fig.1 the front speed versus the time delay is shown. The
continuous line represents the results of the approach proposed in
this paper following Eq.(\ref{eq:Otheresultsev}) and the numerical
simulation by Eq.(\ref{eq:newginzburg}). The results coincides.
Also, the dashed line represents the bound from the variational(BD)
method\cite{artorix}.

\emph{Vector potential $q=1-f$}. For a set of parameters\cite{barto}
$\kappa=1/\sqrt{2}$ and $\bar{\sigma}=1/2$, we have that
$s(z)+n(z)=1$, then Eq.(\ref{eq:steady_equation}) takes the form $
s_{zz}+(c/2)\,s_{z}+\mathfrak{F}(s)=0$, where
$\mathfrak{F}(s)=s^{2}(1-s)$ is the reaction term. Proceeding as in
Eq.(\ref{eq:trii}) we have that,
\begin{eqnarray}
\label{eq:gagain}
g(s)=\left[(1-s\right)/s]^{\alpha^{2}}\,\exp\left(\alpha^{2}/s\right),
\end{eqnarray}
and the velocity is given by,
\begin{eqnarray}
\label{eq:velocinte}
c\simeq\,max_{\alpha\in
\mathbb{D}}\left(\frac{2\,\sqrt{2}}{\alpha}\,\int^{1}_{0}
\mathfrak{F}(s)\,h(s)\,\, ds/\int^{1}_{0} g(s)\,\, ds\right).
\end{eqnarray}
The interface speed is given by,
\begin{eqnarray}
\label{eq:velociv}
c\simeq\,max_{\alpha\in
(0,1)}\left(2\,\sqrt{2}\,\alpha^{3}\,\frac{\Gamma(\alpha^{2})}{\Gamma(1+\alpha^{2})}\right),
\end{eqnarray}
for $\alpha=1/2$ we obtain the maximum for Eq.(\ref{eq:velociv}),
then $c=\sqrt{2}$ which is the result obtained by using the MSH
method.
\section{Front flux expulsion with delay}
It is well known the existence of pinning produces a delay
time\cite{altshuler} in the magnetic field penetration o expulsion.
This can be taken into account by resorting to hyperbolic
differential equations seen in Section I, which generalize the
parabolic equation. The aim of this section is to study of the
interface speed problem in superconducting samples by means of the
HD equations, which can be written as
\begin{eqnarray}
\label{eq:generalHRD}
\tau\,\frac{\partial^2\,u}{\partial\,t^{2}}+\frac{\partial\,u}{\partial\,t}=\,\frac{\partial^2\,u}{\partial\,x^{2}}
+f(u)+\tau\,\frac{\partial\,f(u)}{\partial\,t},
\end{eqnarray}

In the absence of a delay time $(\tau =0)$, this reduces to the
classical equation $u_{t}=u_{xx}+f(u)$.

\emph{Vector potential $q=0$}. Taking into account the
Eqs.(\ref{eq:newginzburg}) and (\ref{eq:generalHRD}) we can write
the following expression,
\begin{eqnarray}
\label{eq:newgeneralHRD}
\kappa^{2}\,\tau\,\frac{\partial^2\,f}{\partial\,t^{2}}+\kappa^{2}\,\frac{\partial\,f}{\partial\,t}=\frac{\partial^2\,f}{\partial\,x^{2}}
+\kappa^{2}\,\mathfrak{F}+\kappa^{2}\,\tau\,\frac{\partial\,\mathfrak{F}}{\partial\,t},
\end{eqnarray}
where $\mathfrak{F} = s(1-s^{2})$.

It has been proved\cite{mendez1,mendez2,mendez3} that
Eq.(\ref{eq:generalHRD}) has traveling wave fronts with profile
$s(x-ct)$ and moving with speed $c>0$. Then we can write
Eq.(\ref{eq:newgeneralHRD}) as follows,
\begin{eqnarray}
\label{eq:transHRDd}
(1-a\,c^{2})\,s_{zz}+c\,[\kappa^{2}-a\,\mathfrak{F}'(s)]\,s_{z}+\mathfrak{F}_{k}(s)=0,
\end{eqnarray}
where $z=x-ct$, $a=\kappa^{2}\,\tau$, $\mathfrak{F}_{k} =
\kappa^{2}\,\mathfrak{F}$, and with boundary conditions
$lim_{z\rightarrow\infty}s=0$,\,$lim_{z\rightarrow-\infty}s=1$, and
$s_{z}<0$ in $(0,1)$; $s_{z}$ vanishes for $z\rightarrow\pm\infty$.

We define $p(s)=-ds/dz>0$ and $g$ such that $h=-dg/ds>0$. Taking
into account $(1-a\,c^{2})hp+(g \mathfrak{F}_{k}/p)\geq
2\sqrt{1-ac^{2}}\,\sqrt{g\,h \,\mathfrak{F}_{k}}$ and following the
BD method we arrive to
\begin{eqnarray}
\label{eq:newveloc}
\frac{c}{\sqrt{1-a\,c^{2}}}\geq\,2\,\kappa\,\frac{\int^{1}_{0}
(g\,h\,\mathfrak{F})^{1/2}\,ds}{\int^{1}_{0}
g(\,\kappa^{2}-a\,\mathfrak{F}')\,ds}.
\end{eqnarray}

In order to obtain the trial function $g(s)$, we take in
consideration that in the lim $s\rightarrow1$ we get
\begin{eqnarray}
\label{eq:lione}
-c\,[\kappa^{2}-a\,\mathfrak{F}'(s)]\,p\,+\,\mathfrak{F}_{\kappa}(s)\simeq0,
\end{eqnarray}
since $s_{zz} \ll s_{z}$. Then, we write an expression for $p$ in
terms of $\mathfrak{F}$,
\begin{eqnarray}
\label{eq:nlione}
p=\mathfrak{F}_{\kappa}/\alpha\,[\kappa^{2}-a\,\mathfrak{F}'(s)].
\end{eqnarray}
The expression for the speed is given by,
\begin{eqnarray}
\label{eq:changeveloc}
\frac{c}{1-a\,c^{2}}\simeq
max_{\alpha\in(0,1)}\,\frac{2\,\kappa^{2}}{\alpha}\,\frac{\int^{1}_{0}
\left[h\,\mathfrak{F}\,/(\,\kappa^{2}-a\,\mathfrak{F}')\right]\,ds}{\int^{1}_{0}
g(\,\kappa^{2}-a\,\mathfrak{F}')\,ds},
\end{eqnarray}
where the integrals can be only solved by numerical methods. Taking
into account Eq.(\ref{eq:nlione}) and the expression\cite{mendez1}
$(1-ac^{2})hp=\mathfrak{F}_{\kappa} g/p$, we have obtained the
relation for the trial function,
\begin{eqnarray}
\label{eq:nglione}
g=\exp\left[-\frac{\alpha^{2}}{(1-ac^{2})}\int\,\frac{(\kappa^{2}-a\,\mathfrak{F}')^{2}}{\mathfrak{F}_\kappa}\,ds\right].
\end{eqnarray}
\begin{figure}[!tbp]
\centerline{
\includegraphics[width=3.2in]{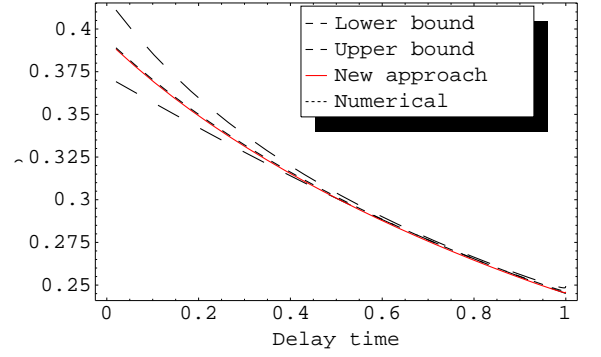}
}\caption{Time-delayed interface propagation speed for the case of
$q=0$ versus the time delay $\tau$.} \label{figure3}
\end{figure}
from where we have for our case,
\begin{eqnarray}
\label{eq:RESULTnglione}
g=\exp\left[\frac{\alpha^{2}}{2\kappa^{2}(1-ac^{2})}\left((3as)^{2}+\lg\left[\frac{(s^{2}-1)^{(2a+\kappa^{2})^{2}}}{s^{2(a-\kappa^{2})^{2}}}\right]\right)\right],
\end{eqnarray}
and for $h$,
\begin{eqnarray}
\label{eq:hexpression}
h(s,\alpha,a)=\frac{\alpha^{2}\,\left(\kappa^{2}+a(3s^{2}-1)\right)^{2}}{\kappa^{2}\,s(s^{2}-1)\,(a\,c^{2}-1)}\,\,g(s,\alpha,\kappa)
\end{eqnarray}

In Fig.2 the front speed versus the time delay is shown. The
continuous line represents the result of the approach proposed in
this paper following Eq.(\ref{eq:changeveloc}) which coincides with
the numerical simulation done using Eq.(\ref{eq:newgeneralHRD}).
Also, we have represented the lower and upper bounds from the
variational(BD) method\cite{artorix}.

\emph{Vector potential $q=1-f$}. Taking into account the
Eqs.(\ref{eq:generalHRD}) and (\ref{eq:newgeneralHRD}) we can write
the following expression,
\begin{eqnarray}
\label{eq:secondgeneralHRD}
\frac{\tau}{2}\,\frac{\partial^2\,f}{\partial\,t^{2}}+\frac{1}{2}\,\frac{\partial\,f}{\partial\,t}=\frac{\partial^2\,f}{\partial\,x^{2}}
+\frac{1}{2}\,\mathfrak{F}+\frac{\tau}{2}\,\frac{\partial\,\mathfrak{F}}{\partial\,t},
\end{eqnarray}
where $\mathfrak{F} = s^{2}(1-s)$.

Then we can write Eq.(\ref{eq:secondgeneralHRD}) as follows,
\begin{eqnarray}
\label{eq:transHRD}
(1-a\,c^{2})\,s_{zz}+c\,[\kappa^{2}-a\,\mathfrak{F}'(s)]\,s_{z}+\mathfrak{F}_{k}(s)=0,
\end{eqnarray}
where we have assumed $\mathfrak{F}_{k}=(1/2)\mathfrak{F}$ and
$a=\tau/2$.

The expression for the velocity is given by
\begin{eqnarray}
\label{eq:newuppers}
\frac{c}{\sqrt{1-ac^{2}}}\geq\,2\,\sqrt{2}\,\frac{\int^{1}_{0}
(g\,h\,\mathfrak{F})^{1/2}\,ds}{\int^{1}_{0}
g(1-2\,a\,\mathfrak{F}')\, ds}.
\end{eqnarray}
In order to obtain an expression for the trial function $g(s)$, we
take in consideration that in the lim $s\rightarrow1$ we get
\begin{eqnarray}
\label{eq:dlione}
-c\,[(1/2)-a\,\mathfrak{F}'(s)]\,p\,+\,\mathfrak{F}_{\kappa}(s)\simeq0,
\end{eqnarray}
since $s_{zz} \ll s_{z}$. Then, we write an expression for $p$ in
terms of $\mathfrak{F}$,
\begin{eqnarray}
\label{eq:Anlione}
p=\mathfrak{F}_{\kappa}/\alpha\,[(1/2)-a\,\mathfrak{F}'].
\end{eqnarray}
The expression for the speed is given by,
\begin{eqnarray}
\label{eq:Nchangeveloc}
\frac{c}{1-a\,c^{2}}\simeq
max_{\alpha\in(0,1)}\,\frac{2\,\kappa^{2}}{\alpha}\,\frac{\int^{1}_{0}
\left[h\,\mathfrak{F}\,/(\,(1/2)-a\,\mathfrak{F}')\right]\,ds}{\int^{1}_{0}
g(\,(1/2)-a\,\mathfrak{F}')\,ds}.
\end{eqnarray}
\begin{figure}[!tbp]
\centerline{
\includegraphics[width=3.2in]{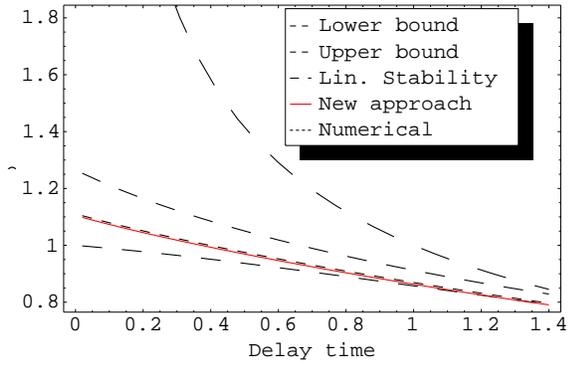}
}\caption{Time-delayed interface propagation speed for $q=1-f$
versus the time delay.} \label{figure3}
\end{figure}
Taking into account Eq.(\ref{eq:Anlione}) and the
expression\cite{mendez1} $(1-ac^{2})hp=\mathfrak{F}_{\kappa} g/p$,
we have obtained the relation for the trial function,
\begin{eqnarray}
\label{eq:Nnglione}
g=\exp\left[-\frac{\alpha^{2}}{(1-ac^{2})}\int\,\frac{((1/2)-a\,\mathfrak{F}')^{2}}{\mathfrak{F}_\kappa}\,ds\right].
\end{eqnarray}
from where we have for our case,
\begin{eqnarray}
\label{eq:NRESULTnglione}
g(s,\alpha,a)\,=\,\exp\left[\frac{\alpha^{2}}{4\,s\,(1-ac^{2})}\left(g_{1}+g_{2}\right)\right],
\end{eqnarray}
where $g_{1}=1+24\,a^{2}\,s^{2}(3s-2)$, and
$g_{2}=\lg\,(s-1)^{s\,(4a+1)^{2}}\,s^{s\,(16\,a-1)}$,
\begin{eqnarray}
\label{eq:Nhexpression}
h(s,\alpha,a)=\frac{\alpha^{2}\,\left[1+4\,a\,s(3s-2)\right]}{4\,s^{2}(1-s)\,(1-a\,c^{2})}\:\,g(s,\alpha,a)
\end{eqnarray}
The integrals in Eq.(\ref{eq:Nchangeveloc}) can be only solved by
numerical methods.

In Fig.3 the front speed versus the time delay is shown. The
continuous line represents the results based on the approach
proposed in this paper following Eq.(\ref{eq:Nchangeveloc}) which
coincides with the numerical simulation done using
Eq.(\ref{eq:secondgeneralHRD}). Also, we have represented the lower
and upper bounds from the variational(BD) method\cite{artorix}.

\emph{Conclusion}. We have computed for the Ginzburg-Landau
equations in the form of parabolic and hyperbolic equations the
superconducting-normal interface propagation speed by a new
approach. This approach is based in the method proposed by Vincent
and Fort in Ref.\onlinecite{vincent}. We have obtained the
expressions for the trial function $g$ in each case developed. The
results of our methodology coincide with the numerical results for
the examples analyzed.

\end{document}